\documentclass[aps,amsmath,showpacs,showkeys,12pt]{revtex4}
\usepackage{pifont}
\usepackage{txfonts}
\usepackage{cases}
\usepackage{dcolumn}
\usepackage{amsmath}
\usepackage{amsfonts}
\usepackage{amssymb}
\usepackage{bm}
\usepackage{graphicx}
\usepackage{subfigure}
\usepackage{psfrag}
\usepackage{wrapfig}
\usepackage{makeidx}
\usepackage{indentfirst}

\begin{document}

\title{Thermodynamics of Third Order Lovelock-Born-Infeld Black
Holes\footnote{This work has been supported by NSFC under grant No. 10875060}}

\author{Peng Li$^{1}$\footnote{Email: lipeng069@gmail.com},
        Rui-Hong Yue$^{1,3}$\footnote{Email: yueruihong@nbu.edu.cn},
        and De-Cheng Zou$^{2}$}
\affiliation{$^{1}$Institute of Modern Physics, Northwest University, Xi'an 710069, China\\
             $^{2}$Department of Physics, Northwest University, Xi'an 710069, China\\
             $^{3}$Faculty of Science, Ningbo University, Ningbo 315211, China}

\begin{abstract}

We here explore black holes in the third order Lovelock
gravity coupling with nonlinear Born-Infeld electromagnetic field.
Considering special second and third order coefficients ($\hat{\alpha}_2^2=3\hat{\alpha}_3=\alpha^2$),
we analyze the thermodynamics of third order Lovelock-Born-Infeld black holes
and, in 7-dimensional AdS space-time, discuss the stability of black holes in
different event horizon structures. We find that the
cosmological constant $\Lambda$ plays an important role in
the distribution of stable regions of black holes.

\end{abstract}

\pacs{04.70.Dy, 04.50.-h}

\keywords{Third order Lovelock gravity; Born-Infeld; Thermodynamics; Stability.}

\maketitle

\section{Introduction\label{Introduction}}

In higher dimensional space-time, the Lovelock gravity involving higher-derivative curvature terms, as a most natural
generalization of Einstein gravity, only contains the terms without more than second order derivatives of metric
\cite{Lovelock:1971yv}.
It may appear in a quantum gravity theory and the low-energy effective action of string theory, and
 has been found to be free of ghost when expanding on a flat
space\cite{Boulware:1985wk, Witten:1998qj}. Due to the AdS/CFT correspondence,
these extended terms give a correction to the large N expansion of boundary CFTs on
the side of dual field theory\cite{Cai:2001dz, Cai:2003gr, Cai:2003kt}.
So far, the exact static and spherically symmetric black hole
solutions have been investigated in \cite{Dehghani:2009zzb}, thermodynamics of black holes
in \cite{Ge:2009ac, Dehghani:2009zzb, Dehghani:2005zzb, Zou:2010yr} in third order Lovelock gravity.
In addition, some rotating black branes \cite{Dehghani2006cu, Dehghani2006cp, Hendi2010cp}
and slowly rotating black holes \cite{Kim:2007iw, Zou:2010dx, Yue:2011zz, Yue:2011et}
have been discussed in the second (Gauss-Bonnet) and third order Lovelock gravity.

Besides the curvature terms, the  higher derivative gauge field may also
contribute to Lovelock gravity. This is done by explicitly constructing black holes
solutions coupled to a Born-Infeld (BI) gauge field in the presence of a
cosmological constant. Its Lagrangian ${\cal L(F)}$ is given by
\begin{eqnarray}
{\cal L(F)}=4\beta^2(1-\sqrt{1+\frac{F^{\mu\nu}F_{\mu\nu}}{2\beta^2}}),\label{eq:4a}
\end{eqnarray}
where $\beta$ is the BI parameter,
$F_{\mu\nu}=\partial_{\mu}A_{\nu}-\partial_{\nu}A_{\mu}$ is electromagnetic tensor
field with vector potential $A_{\mu}$. The BI theory was originally
introduced to get a classical theory of charged particles with finite
self-energy \cite{Born:1934gh}. Hoffmann \cite{Hoffmann:1935ty}
related the BI electromagnetic
field with general relativity. The BI black hole with zero cosological constant was obtained by Garcia $et$
$al$ \cite{Garica:1984}. Later, the spherically symmetric Einstein-Born-Infeld black
hole solutions with cosmological constant were studied in \cite{Cai:2004eh, Fernando:2003tz, Dey:2004yt},
Born-Infeld-dilaton models in \cite{Sheykhi:2008rt, Sheykhi:2007gw, Sheykhi:2006dz}.
For the Lovelock gravity, the Gauss-Bonnet-Born-Infeld black hole solutions were
found in asymptotically flat \cite{Wiltshire:1988uq},
and anti-de Sitter space-time \cite{Aiello:2004rz, Zou:2010tv}. Moreover,
the third order Lovelock-Born-Infeld black hole solutions with special coefficients
have been also obtained in anti-de Sitter space-time \cite{Dehghani:2008qr}.
In this paper, we are going to analyze the black holes with two independent Gauss-Bonnet
and third order Lovelock coefficients in third order Lovelock-Born-Infeld gravity.
Then, thermodynamics of black holes with special second and third order coefficients
($\hat{\alpha}_2^2=3\hat{\alpha}_3=\alpha^2$) will be discussed in different
event horizon structures $k=0, \pm1$.

The outline of present paper is as follows. We investigate black holes in third order
Lovelock-Born-Infeld gravity in Sec.~\ref{Solution}. Then, the stability of black holes
in 7-dimensional AdS space-time will be discussed in Sec.~\ref{Stability}.
Finally, we end our paper with closing remarks in Sec.~\ref{Conclusions}.

\section{Lovelock-Born-Infeld Black Holes\label{Solution}}

Coupling with a nonlinear BI filed, in $(n+1)$ dimensions, the action of third order
Lovelock gravity can be given as
\begin{equation}
I=\frac{1}{16\pi}\int{d^{n+1}x \sqrt{-g}\left[-2\Lambda+\mathcal{L}_1+\alpha_2\mathcal{L}_2
+\alpha_3\mathcal{L}_3+L(F)\right]}\label{action}
\end{equation}
where
\begin{eqnarray}
\mathcal{L}_1&=&R \\
\mathcal{L}_{2}&=&R_{\mu\nu\rho\sigma}R^{\mu\nu\rho\sigma}-4R_{\mu\nu}R^{\mu\nu}+R^2,\label{GB term}\\
\mathcal{L}_3&=&2R^{\mu\nu\sigma\kappa}R_{\sigma\kappa\rho\tau}{R^{\rho\tau}}_{\mu\nu}
                +8{R^{\mu\nu}}_{\sigma\rho}{R^{\sigma\kappa}}_{\nu\tau}{R^{\rho\tau}}_{\mu\kappa}
                +24R^{\mu\nu\sigma\kappa}R_{\sigma\kappa\nu\rho}{R^{\rho}}_{\mu}
                +3RR^{\mu\nu\sigma\kappa}R_{\sigma\kappa\mu\nu}\nonumber\\
              &&+24R^{\mu\nu\sigma\kappa}R_{\sigma\mu}R_{\kappa\nu}+16R^{\mu\nu}R_{\nu\sigma}{R^{\sigma}}_{\mu}
                -12RR^{\mu\nu}R_{\mu\nu}+R^3.
\end{eqnarray}
$\alpha_2$ and $\alpha_3$ here are the second and third order Lovelock coefficients
and the cosmological constant reads $\Lambda$.

In order to obtain  black hole solutions, we assume the static and spherically symmetric metric to be
\begin{eqnarray}
ds^2=-f(r)dt^2+\frac{1}{f(r)}dr^2+r^2d\Sigma_{_{n-1}}^{2},\label{metric}
\end{eqnarray}
where $d\Sigma_{n-1}^{2}$ denotes the line element of an $n-1$-dimensional space
with constant curvature $(n-1)(n-2)k$
\begin{equation}
d\Sigma_{_{n-1}}^{2}=\left\{
    \begin{aligned}
    &d{\theta_1}^2+\sum_{i=2}^{n-1}{\prod_{j=1}^{i-1}{\sin^2\theta_j d{\theta_i}^2}}& &k=1,\\
    &d{\theta_1}^2+\sinh^2\theta_1 d{\theta_2}^2+\sinh^2\theta_1 \sum_{i
    =3}^{n-1}{\prod_{j=2}^{i-1}{\sin^2\theta_j d{\theta_i}^2}}& &k=-1\\
    &\sum_{i=1}^{n-1}{d{\phi_i}^2}& &k=0.
    \end{aligned} \right.,\label{Omega-n-1}
\end{equation}

Substituting the metric Eq.~\eqref{metric} into the action Eq.~\eqref{action},
then varying it with respect to the metric tensor $g_{\mu\nu}$
and electromagnetic vector filed $A_{\mu\nu}$, the gravitational
and electromagnetic field equation can be written as:
\begin{equation}
-\frac{2\Lambda r^{n-1}}{n-1}+(r^n\varphi+\hat{\alpha}_2r^n\varphi^2+\hat{\alpha}_3r^n\varphi^3)~'+\frac{4\beta^2 r^{n-1}}{(n-1)}(1-\frac{1}{\sqrt{1+\frac{F^2}{2\beta^2}}})=0,\label{gra-fun1}
\end{equation}
\begin{equation}
\partial_\mu \left(\frac{\sqrt{-g}F^{\mu\nu}}{\sqrt{1+\frac{F^2}{2\beta^2}}}\right)=0.\label{eletromagnetic}
\end{equation}
Here, the function $\varphi(r)$ represents $\frac{k-f(r)}{r^2}$ and coefficients read
\begin{eqnarray}
\hat{\alpha}_2=\alpha_2 (n-2)(n-3),\quad
\hat{\alpha}_3=\alpha_3 (n-2)(n-3)(n-4)(n-5)\nonumber.
\end{eqnarray}

First of all, the electromagnetic field equation Eq.\eqref{eletromagnetic} can be
solved in a gauge with only one novanishing component $F^{rt}$
\begin{equation}
F^{rt}=\frac{\sqrt{(n-1)(n-2)}\beta q}{\sqrt{2\beta^2r^{2(n-1)}+(n-1)(n-2)q^2}},\label{Frt}
\end{equation}
where an integration constant $q$ denotes the electric charge of black holes and
the electromagnetic field is infinite at $r=0$.
The corresponding magnetic vector potential can be expressed as
\begin{equation*}
\begin{array}{rcl}
A_\mu&=&-\sqrt{\frac{n-1}{2(n-2)}}\frac{q}{r^{n-2}}\digamma(\eta)\delta_{\mu}^{0},\label{magnetic potential}\\[5mm]
\digamma(\eta)&=& _2F_{1}\left(\left[\frac{1}{2},\frac{n-2}{2(n-1)}\right],
\left[\frac{3n-4}{2(n-1)}\right],-\eta\right),\quad \eta=\frac{(n-1)(n-2)q^2}{2\beta^2 r^{2n-2}}.
\end{array}
\end{equation*}
Notice that $\digamma(\eta)$ is a hypergeometric function and comes from the integration equation: $_2F_{1}\left(\left[\frac{1}{2},\frac{n-2}{2(n-1)}\right],\left[\frac{3n-4}{2(n-1)}\right], bu^{2n-2}\right)
 =\frac{n-2}{u^{n-2}}\int{\frac{u^n-3}{\sqrt{1-bu^{2n-2}}}du}$.
 In the case of $\eta \rightarrow 0$ $(\beta \rightarrow \infty)$, we get $\digamma(\eta) \rightarrow 1$
and causes that $A_\mu$ reduces to the standard form of Maxwell field.

Turning to the gravitational equation, one can easily acquire by substituting Eq.~\eqref{Frt} into Eq.~\eqref{gra-fun1}
\begin{eqnarray}
{\hat{\alpha}}_3\varphi^3+{\hat{\alpha}}_2\varphi^2+\varphi
        -\frac{\omega}{r^n}+\frac{4\beta^2}{n(n-1)}\left[1-\sqrt{1+\eta}-\frac{\Lambda}{2\beta^2}
                                                    +\frac{(n-1)\eta}{(n-2)}\digamma(\eta)\right]=0,\label{gra-fun2}
\end{eqnarray}
where $\omega$ is an integration constant relating to the ADM mass of black holes.
By solving the cubic equation, the general solution of Eq.~\eqref{gra-fun2} is
\begin{eqnarray*}
f(r)&=&k-r^2\varphi(r)=k+\frac{{r^2\hat{\alpha}}_2}{3{\hat{\alpha}}_3}
    +\frac{2^{1/3}A r^2}{3{\hat{\alpha}}_3\left[B+\sqrt{4A^3+B^2}\right]^{1/3}}
    -\frac{r^2}{3\cdot2^{1/3}{\hat{\alpha}}_3}\left[B+\sqrt{4A^3+B^2}\right]^{1/3}, \\
A&=&-{\hat{\alpha}}_2^2+3{\hat{\alpha}}_3,\nonumber\\
B&=&-2{\hat{\alpha}}_2^3+9{\hat{\alpha}}_2{\hat{\alpha}}_3-27{\hat{\alpha}}_3^2
                 \left[-\frac{\omega}{r^n}+\frac{4\beta^2}{n(n-1)}\left(1-\sqrt{1+\eta}-\frac{\Lambda}{2\beta^2}
                    -\frac{(n-1)\eta}{(n-2)}\digamma(\eta)\right)\right].
\end{eqnarray*}
If $\hat{\alpha}_2^2=3\hat{\alpha}_3=\alpha^2$, it can be reduced to\cite{Dehghani:2008qr}
\begin{eqnarray}
f(r)&=&k+\frac{r^2}{\alpha}(1-g(r)^{\frac{1}{3}}),\nonumber\\
g(r)&=&1+\frac{3\alpha \omega}{r^n}
    -\frac{12\alpha\beta^2}{n(n-1)}\left[1-\sqrt{1+\eta}
    -\frac{\Lambda}{2\beta^2}+\frac{(n-1)\eta}{(n-2)}\digamma(\eta) \right].\label{fr}
\end{eqnarray}

Based on the black hole solution, the thermodynamics of black holes will be investigated.
First, the Arnowitt-Deser-Misner (ADM) mass of black holes is
\begin{eqnarray}
M&=&\frac{(n-1)V_{n-1}}{16\pi}\omega\nonumber\\
 &=&\frac{(n-1)V_{n-1}}{16\pi}
            \left[k\left({r_+}^{n-2}+k\alpha{r_+}^{n-4}+\frac{1}{3}k^2\alpha^2 {r_+}^{n-6}\right)\right.\nonumber\\
 & &\left.+\frac{4\beta^2{r_+}^n}{n(n-1)}\left(1-\sqrt{1+\eta_+}-\frac{\Lambda}{2\beta^2}
            +\frac{(n-1)\eta_+}{(n-2)}\digamma(\eta_+)\right)\right],\label{mass}
\end{eqnarray}
where, $\omega$ is related to $f(r)=0$ at horizon radius $r=r_+$ in Eq.~\eqref{fr}.
The Hawking temperature of black holes on the outer horizon radius $r_+$
can be easily obtained by \cite{Dehghani:2008qr}
\begin{eqnarray}
T_+&=&\frac{f'(r_+)}{4\pi}\nonumber\\
   &=&\frac{(n-1)k\left[3(n-2)r_+^4+3(n-4)k\alpha r_+^2+(n-6)k^2\alpha^2\right]
   +12\beta^2r_+^6(1-\sqrt{1+\eta_+})-6\Lambda r_+^6}{12\pi(n-1)r_+({r_+}^2+k\alpha)^2}\label{temperature}
\end{eqnarray}
and the entropy of black holes is
\begin{equation}
S=\frac{V_{n-1}}{4}\left[r_+^4+\frac{2(n-1)}{n-3}\alpha r_+^2+\frac{n-1}{n-5}\alpha^2\right]r_+^{n-5}.\label{entropy}
\end{equation}

Moreover, the charge of black holes can be calculated by integrating the flux
of the electric field at infinity\cite{Cai:2004eh}
\begin{equation}
Q=\frac{1}{4\pi}\int^*{F d\Omega}
 =\frac{V_{n-1}}{4\pi}\sqrt{\frac{(n-1)(n-2)}{2}} q.\label{charge}
\end{equation}
The electric potential $\Phi$ at infinity with respect to the horizon is defined by
\begin{equation}
\Phi=A_{\mu}\chi^{\mu}|_{r \rightarrow \infty}-A_{\mu}\chi^{\mu}|_{r=r_+}
    =\sqrt{\frac{n-1}{2(n-2)}}\frac{q}{r_+^{n-2}}\digamma({\eta_+}).\label{electric potential}
\end{equation}

\section{Stability of Black Holes\label{Stability}}

In order to perform the stability analysis of black holes, the heat capacity of black holes
is given as
\begin{equation}
C_Q=\left(\frac{\partial M}{\partial T}\right)_Q
   =\left(\frac{\partial M}{\partial r_+}\right)_Q~ \Big/ ~\left(\frac{\partial T_+}{\partial r_+}\right)_Q,\label{CQ}
\end{equation}
where
\begin{eqnarray*}
\left(\frac{\partial M}{\partial r_+}\right)_Q
&=&\frac{1}{4}(n-1)V_{n-1}{r_+}^{n-6}({r_+}^2+k\alpha)^2T_+,\\
\left(\frac{\partial T_+}{\partial r_+}\right)_Q
&=&\frac{1}{12\pi(n-1){r_+}^2({r_+}^2+k\alpha)^3}
\left\{(n-1)k\left[-3(n-2){r_+}^6+18k\alpha{r_+}^4-2(n-9)k^2\alpha^2{r_+}^2\right.\right.\\
& &\left.-(n-6)k^3\alpha^3\right]+12\beta^2{r_+}^6({r_+}^2+5k\alpha)+12\beta^2{r_+}^6
                                    \left[(n-2){r_+}^2+(n-6)k\alpha\right]\sqrt{1+\eta_+}\\
& &-12(n-1)\beta^2{r_+}^6\frac{{r_+}^2+k\alpha}{\sqrt{1+\eta_+}}-6r_+^6 \Lambda({r_+}^2+5k\alpha)\Big\}.
\end{eqnarray*}

It is obvious that the heat capacity depends on the coefficient $\alpha$ of Lovelock term,
Born-Infeld parameter $\beta$, event structure $k$, cosmological constant $\Lambda$ and
dimension $(n+1)$. When $\alpha\rightarrow 0$, it returns to the Einstein-Born-Infeld case.
In the following part, we are going to analyze the stability of the Lovelock-Born-Infeld
black holes in which the cosmological constant is defined as $\Lambda=-\frac{n(n-1)}{2l^2}$.

\subsection{The case $k=0$}

For $k=0$, the mass and temperature of black holes are simplified as
\begin{eqnarray*}
M&=&\frac{V_{n-1}\beta^2{r_+}^n}{4n\pi}\left[1-\sqrt{1+\eta_+}-\frac{n(n-1)}{4\beta^2l^2}
            +\frac{(n-1)\eta_+}{(n-2)}\digamma(\eta_+)\right],\\
T&=&\frac{2\beta^2r_+^3(1-\sqrt{1+\eta_+})+n(n-1)r_+^3/2l^2}{2\pi(n-1)},
\end{eqnarray*}
then, one can find the extremal mass by calculating $r_+$ from $T(r_+)=0$
\begin{eqnarray}
m_{ext}=\frac{2(n-1)q_{ext}^2}{n}\left(\frac{-\frac{n(n-1)}{2l^2}(-\frac{n(n-1)}{2l^2}
-4\beta^2)}{2(n-1)(n-2)\beta^2q_{ext}^2}\right)^{\frac{n-2}{2n-2}}
\digamma\left(\frac{-\frac{n(n-1)}{2l^2}(-\frac{n(n-1)}{2l^2}-4\beta^2)}{4\beta^4}\right).\label{extremal mass}
\end{eqnarray}
It is shown that, if the mass $m>m_{ext}$, the black hole exists with minimum horizon $r_{ext}$,
that is to say, it gives the minimum mass of this kind of black holes. The heat capacity can be given as
\begin{eqnarray}
C_Q=\frac{(n-1)V_{n-1}{r_+}^{n-1}\left[2\beta^2(1-\sqrt{1+\eta_+})+\frac{n(n-1)}{2l^2}\right]}
    {4\left[2\beta^2+2(n-2)\beta^2\sqrt{1+\eta_+}
        -2(n-1)\beta^2\frac{1}{\sqrt{1+\eta_+}}+\frac{n(n-1)}{2l^2}\right]},\label{CQk0}
\end{eqnarray}
where $V_{n-1}=2\pi^n/\Gamma(\frac{n}{2})$. In the case $k=0$,
the heat capacity does not contain the coefficient $\alpha$. That means Lovelock
correction has no contribution in this condition. If defined the dimension as $n+1=4$,
the heat capacity is correctly equal to the Einstein-Born-Infeld black hole\cite{Myung:2008eb}.

\begin{figure}
  \subfigure[~$\beta=0.1$, $l=1$ and $q=1$]{
    \label{CQ-k0-beta}
    \includegraphics{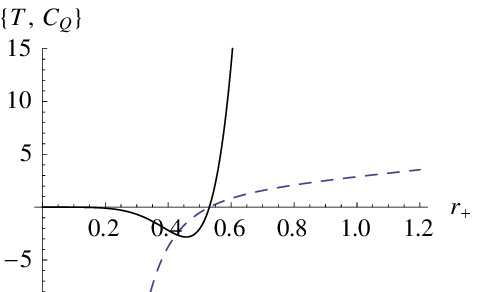}}
  \subfigure[~$\beta=0.1$, $l=1$ and $q=0$]{
    \label{CQ-k0-q}
    \includegraphics{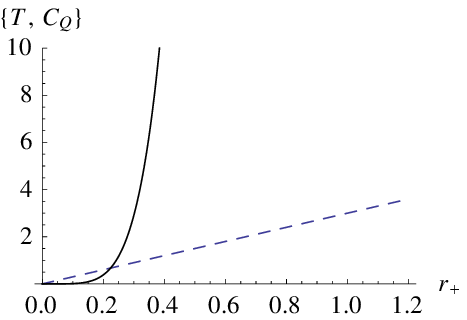}}
  \subfigure[~$\beta=0.1$, $l\rightarrow \infty$ and $q=1$]{
    \label{CQ-k0-l}
    \includegraphics{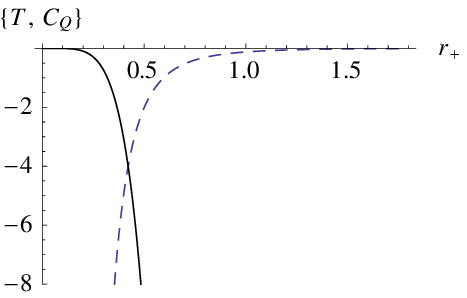}}
  \caption{Temperature $T$ and heat capacity $C_Q$ vs. horizon radius $r_+$
  in 7-dimensional space-time for $k=0$.}
  \label{CQ-k0}
\end{figure}

In 7-dimensional AdS space-time, we plot the temperature $T$ and heat capacity $C_Q$ as
a function of horizon radius $r_+$ with different parameter values in Fig.~\ref{CQ-k0}.
When $q\neq 0$ and $\Lambda\neq 0$, the temperature $T$ disappears at $r_+=r_{ext}$
and increases as the horizon radius $r_+ \rightarrow \infty$. In addition,
the heat capacity $C_Q$ decreases accelerando from zero to reach the negative extremal pole,
and then increases rapidly to approach the positive infinity as $r_+ \rightarrow \infty$.
Thus, the black holes are locally stable for $r_+>r_{ext}$.
Moreover, the case of $q=0$ shows that the heat capacity $C_Q$
is always positive\subref{2} and then the uncharged third order Lovelock black holes
are always locally stable in the whole range of $r_+$.
While the temperature $T$ and heat capacities $C_Q$ always maintains negative in the
case of $\Lambda=0(l=\infty)$, that is to say, the black holes are thermodynamically unstable
in the whole range for $\Lambda=0$ and stable in AdS space-time in the region of $r_+>r_{ext}$.

\subsection{The case $k=1$}

\begin{figure}
  \subfigure[~$\beta=0.5$, $l=5$ and $q=1$]{
    \label{T-k1-alpha}
    \includegraphics{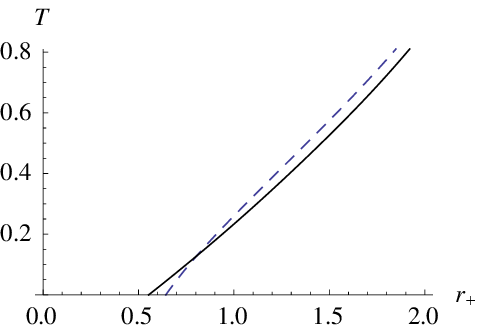}}
  \subfigure[~$\alpha=0.5$, $l=5$ and $q=1$]{
    \label{T-k1-beta}
    \includegraphics{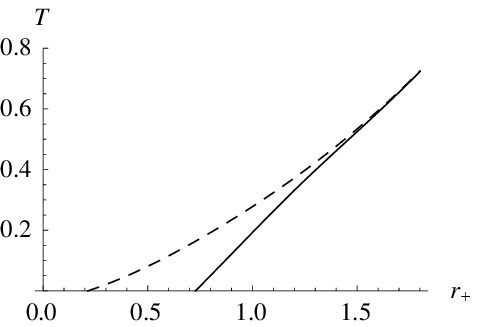}}
  \subfigure[~$\alpha=0.5$ and $\beta=0.5$]{
    \label{T-k1-sl}
    \includegraphics{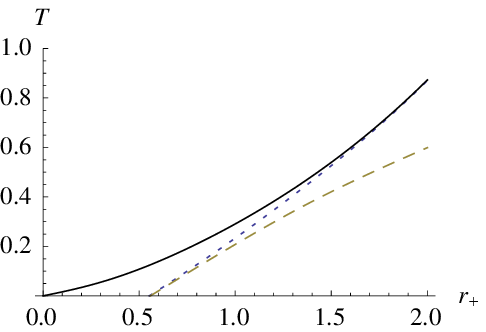}}
  \subfigure[~$\beta=0.5$, $l=5$, $q=1$]{
    \label{M-k1-alpha}
    \includegraphics{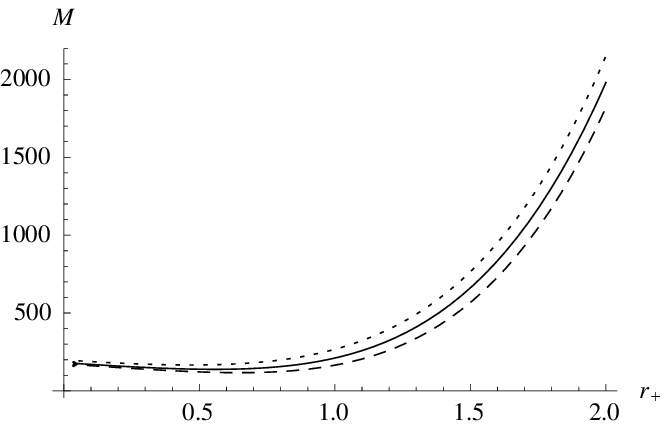}}
  \subfigure[~$\alpha=0.5$, $l=5$, $q=1$]{
    \label{M-k1-beta}
    \includegraphics{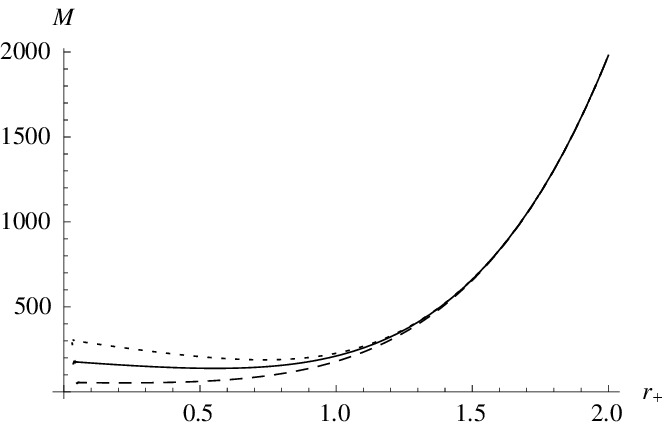}}
  \caption{T and M vs. $r_+$ for $k=1$:\subref{T-k1-alpha}\subref{M-k1-alpha}$\alpha$=0.1(Dashed line), 0.5(Solid line), 0.9(Dotted line);
  \subref{T-k1-beta}\subref{M-k1-beta} $\beta$=0.1(dashed line), 1.0(solid line);
  \subref{T-k1-sl}l=5 and q=1(dotted line), l=5 and q=0(solid line),
  $l\rightarrow \infty$ and q=1 (dashed line).}\label{TM-k1}
\end{figure}

It is quite difficult to calculate the extremal case of black holes for $k=1$.
Using Eq.~\eqref{mass}\eqref{temperature}, we plot the temperature and mass versus horizon radius
in Fig.~\ref{TM-k1}. Obviously, the temperature $T$ of black holes is zero at
a certain horizon radius $r_+$ where mass has the minimum value.
It means that the extremal black holes exist
in asymptotically flat and AdS space-time and vanish for the uncharged black holes.

\begin{figure}
  \subfigure[~$\beta=0.5$, $l=5$ and $q=1$]{
    \label{CQ-k1-alpha}
    \includegraphics{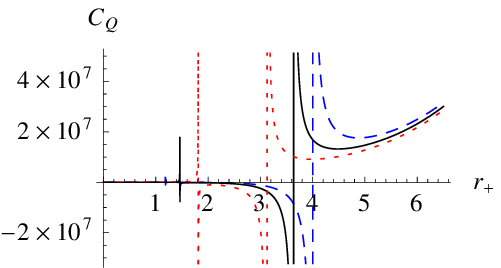}}
  \subfigure[zoom in at (0,1)]{
    \label{micro1}
    \includegraphics{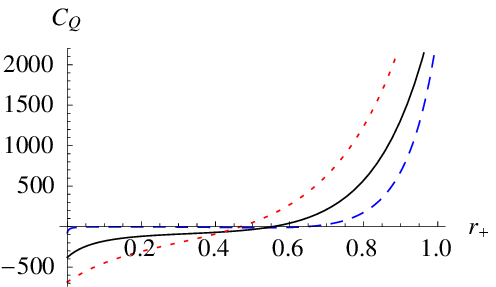}}
  \subfigure[zoom in at (1,2)]{
    \label{micro2}
    \includegraphics{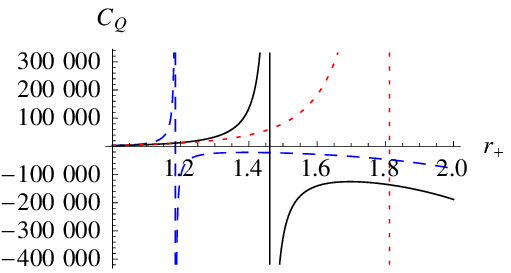}}
  \caption{$C_Q$ vs. $r_+$ for $k=1$, $\alpha$=0.1(dashed line), 0.5(solid line)
  and 0.9(dotted line).}\label{CQ-k1}
\end{figure}

\begin{figure}
 \subfigure[~$\alpha=0.5$ and $\beta=0.5$]{
    \label{CQ-k1-sl}
    \includegraphics{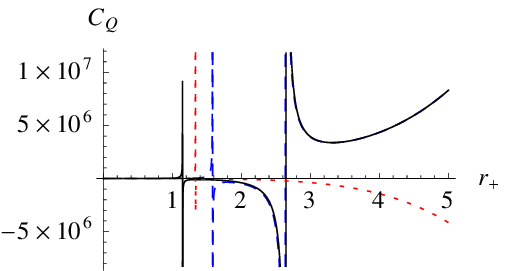}}
  \subfigure[~zoom in at (0,1)]{
    \label{CQ-k1-s2}
    \includegraphics{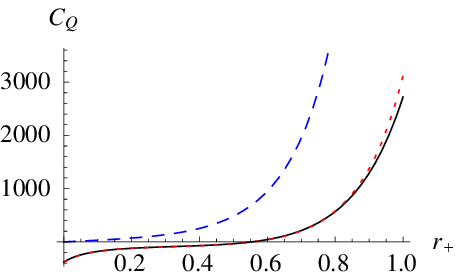}}
  \subfigure[~zoom in at (1,2.8)]{
    \label{CQ-k1-s3}
    \includegraphics{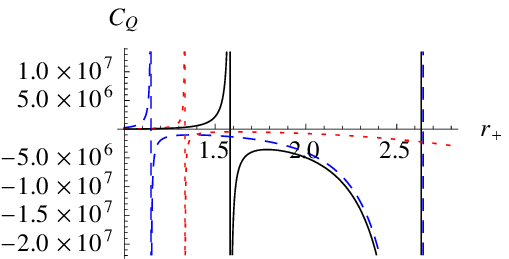}}
  \caption{$C_{Q}$ vs. $r_+$ for $k=1$, l=4 and $q$=0(dashed line), q=1 and $l\rightarrow \infty$(dotted line),
  q=1 and l=4(solid line).}\label{CQ-k1-l}
\end{figure}

The heat capacity $C_Q$ for different parameters are given in Fig.~\ref{CQ-k1}.
With regard to the case of $q\neq 0$ and $\Lambda\neq 0$,
the heat capacity $C_Q$ vanishes at $r_+=r_{ext}$ corresponding to $T=0$.
Later, it blows up at $r_{C1}$ and changes sign. Finally, it blows up
at $r_{C2}$ again and then becomes positive.
Therefore, there exist an intermediate unstable phase for black holes.
Furthermore, we also take into account the effect of gravitational
background in Figs.~\ref{CQ-k1-l}. The heat capacity $C_Q$ is always positive and then
the uncharged black holes are locally stable in the whole range of $r_+$.
While the Dotted line in this figure shows the heat capacity of black holes
in asymptotically flat space-time maintains positive is a small range of $r_+$,
and then there exist an intermediate stable phase.
Obviously, the cosmological constant $\Lambda$ plays an important role in
the distribution of stable regions of black holes.

\subsection{The case $k=-1$}

\begin{figure}
  \subfigure[~$\beta=0.5$, $l=5$, $q=1$]{
    \label{T-k-1-alpha}
    \includegraphics{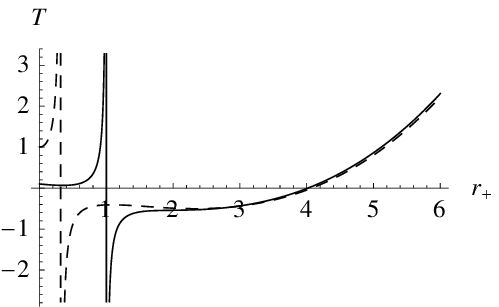}}
  \subfigure[~$\alpha=0.5$, $l=5$, $q=1$]{
    \label{T-k-1-beta}
    \includegraphics{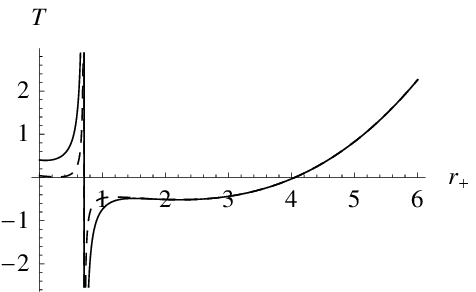}}
  \subfigure[~$\alpha=0.5$ and $\beta=0.5$]{
    \label{T-k-1-sl}
    \includegraphics{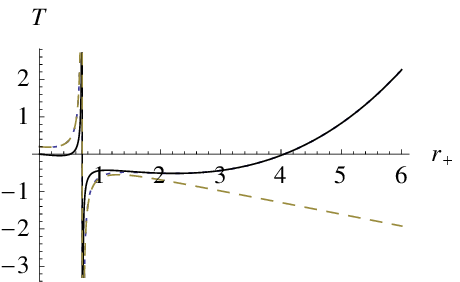}}
  \subfigure[~$\beta=0.5$, $l=5$, $q=1$]{
    \label{M-k-1-alpha}
    \includegraphics[width=0.38\textwidth]{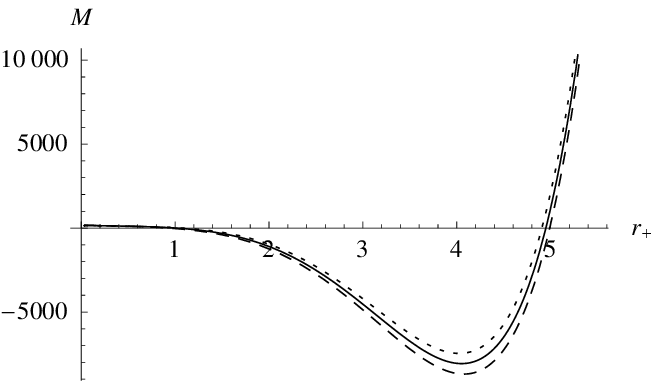}}
  \subfigure[~$\alpha=0.5$, $l=5$, $q=1$]{
    \label{M-k-1-beta}
    \includegraphics[width=0.38\textwidth]{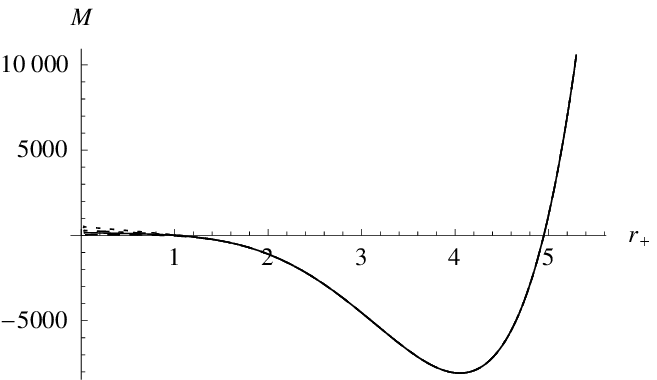}}
  \caption{T and M vs. $r_+$ for $k=-1$: \subref{T-k-1-alpha}\subref{M-k-1-alpha}$\alpha$=0.1(Dashed line), 1(Solid line);
  \subref{T-k-1-beta}\subref{M-k-1-beta}$\beta$=0.1(Dashed line), 1(Solid line);
   \subref{T-k-1-sl}l=5 and q=1(Dotted line), l=5 and q=0(Solid line),
  $l\rightarrow \infty$ and q=1(Dashed line).}\label{TM-k-1}
\end{figure}

We here discuss stability in the case $k=-1$. From Fig.~\ref{TM-k-1},
one can find that, in AdS space-times, the temperature $T$ of black holes is positive when  $r_{ext}<r_+<\infty$,
and will change the sign at $r_+=r_{ext}$. The figure also shows the existence of minimal black
hole with positive temperature, which is in the inside of a singularity. This phenomena also appears in Gauss-Bonnet
gravity \cite{Cai:2001dz}. Considering the $r_+$ dependence of ADM mass,
one can find that the black hole with mass locates in $r_{ext}<r_+<\infty$. Thus, the black hole solution with $r_+<r_{ext}$ is un-physical,
and physical black hole should satisfy $r_{ext}<r_+$.

\begin{figure}
  \subfigure[~$\beta=0.5$, $l=5$, $q=1$]{
    \label{CQ-k-1-alpha}
    \includegraphics{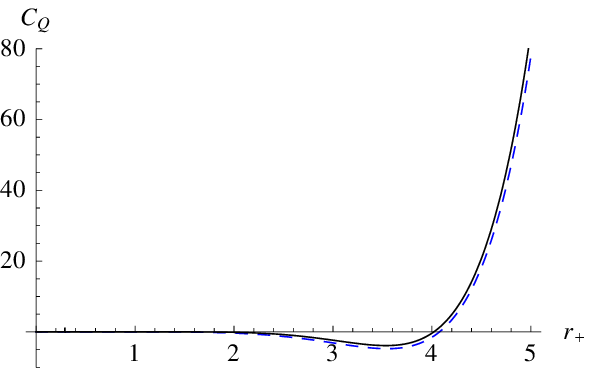}}
  \subfigure[~$\alpha=0.5$, $l=5$, $q=1$]{
    \label{CQ-k-1-beta}
    \includegraphics{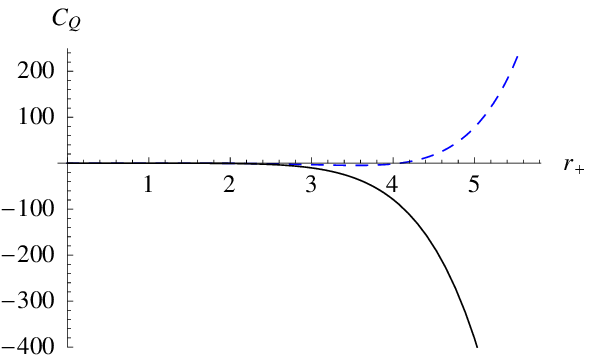}}
  \caption{$C_Q$ vs. $r_+$ for $k=-1$:
  \subref{CQ-k-1-alpha}$\alpha$=0.1(dashed line), 1(solid line);
  \subref{CQ-k-1-beta} l=5 and q=0(dashed line), $l\rightarrow \infty$ and q=1(solid line).}\label{CQ-k-1}
\end{figure}

For the different $\alpha$ and $\beta$, the heat capacities as a function of horizon radius
are plotted in Fig.~\ref{CQ-k-1}.
The heat capacity $C_Q$ also vanishes at the radius $r_{ext}$ for $T=0$ and then
reaches positive infinity as $r_+\rightarrow \infty$. Therefore, the black holes with negative
constant curvature hypersurface horizon in AdS space-time are thermodynamically stable
in the range of $r_{ext}<r_+<\infty$. The Fig.~6(b) also demonstrates
heat capacity $C_Q>0$ in the domain $0<r_+<r_{ext}$. It shows that the
asymptotically flat black holes are thermodynamically stable in this domain.

\section{closing remarks\label{Conclusions}}

We have presented the black hole solutions and discussed the thermodynamics of black holes in third
order Lovelock-Born-Infeld gravity. Considering $\hat{\alpha}_2^2=3\hat{\alpha}_3=\alpha^2$,
the extremal black holes exist for the three cases $k=0, \pm1$ in AdS space-time .

For $k=0$, the black holes are thermodynamically stable in the region $r_+>r_{ext}$ in AdS space-time,
while asymptotically flat black holes are unstable in the whole rang of horizon radius.
With positive constant curvature hypersurface horizon, there exist an
intermediate unstable phase for the AdS black holes and stable phase for asymptotically black holes.
Furthermore, the AdS black holes are locally stable in the domain of $r_+>r_{ext}$ for $k=-1$.
However, the flat black holes are only stable in a small range of $r_+$.

{\bf Acknowledgements}

P.~Li would like to thank Faculty of Science, Ningbo University for the hospitality,
and helpful discussions with our team members.

\end{document}